# Deep learning assessment of breast terminal duct lobular unit involution: towards automated prediction of breast cancer risk

Suzanne C Wetstein[1*], Allison M Onken[2], Christina Luffman[2], Gabrielle M Baker[2], Michael E Pyle[2], Kevin H Kensler[3], Ying Liu[4], Bart Bakker[5], Ruud Vlutters[5], Marinus B van Leeuwen[5], Laura C Collins[2], Stuart J Schnitt[6], Josien PW Pluim[1], Rulla M Tamimi[7], Yujing J Heng[2,#] and Mitko Veta[1,#]

[1] Medical Image Analysis Group, Department of Biomedical Engineering, Eindhoven University of Technology, Eindhoven, The Netherlands

[2] Department of Pathology, Harvard Medical School, Beth Israel Deaconess Medical Center, Boston, MA, USA

[3] Division of Population Sciences, Dana Farber Cancer Institute, Boston, MA, USA

[4] Division of Public Health Sciences, Department of Surgery, Washington University School of Medicine and Alvin J. Siteman Cancer Center, St Louis, MO, USA

[5] Philips Research Europe, High Tech Campus, Eindhoven, The Netherlands

[6] Dana-Farber/Brigham and Women's Cancer Center, Harvard Medical School, Dana-Farber Cancer Institute-Brigham and Women's Hospital, Boston, MA, USA

[7] Channing Division of Network Medicine, Department of Medicine, Harvard Medical School, Brigham and Women's Hospital, Boston, MA, USA

# co-senior authors

*Corresponding Author: Suzanne Wetstein, Medical Image Analysis Group, Department of Biomedical Engineering, Groene Loper 5, 5612AE Eindhoven, the Netherlands; Email: s.c.wetstein@tue.nl; Phone: +31 40-247 5581

## Abstract

Terminal ductal lobular unit (TDLU) involution is the regression of milk-producing structures in the breast. Women with less TDLU involution are more likely to develop breast cancer. A major bottleneck in studying TDLU involution in large cohort studies is the need for labor-intensive manual assessment of TDLUs. We developed a computational pathology solution to automatically capture TDLU involution measures.

Whole slide images (WSIs) of benign breast biopsies were obtained from the Nurses' Health Study (NHS). A first set of 92 WSIs was annotated for TDLUs, acini and adipose tissue to train deep convolutional neural network (CNN) models for detection of acini, and segmentation of TDLUs and adipose tissue. These networks were integrated into a single computational method to capture TDLU involution measures including number of TDLUs per tissue area (mm$^2$), median TDLU span (µm) and median number of acini per TDLU. We validated our method on 40 additional WSIs by comparing with manually acquired measures.

Our CNN models detected acini with an F1 score of 0.73±0.09, and segmented TDLUs and adipose tissue with Dice scores of 0.86±0.11 and 0.86±0.04, respectively. The inter-observer ICC scores for manual assessments on 40 WSIs of number of TDLUs per tissue area, median TDLU span, and median acini count per TDLU were 0.71, 95% CI [0.51, 0.83], 0.81, 95% CI [0.67, 0.90], and 0.73, 95% CI [0.54, 0.85], respectively. Intra-observer reliability was evaluated on 10/40 WSIs with ICC scores of >0.8. Inter-observer ICC scores between automated results and the mean of the two observers were: 0.80, 95% CI [0.63, 0.90] for number of TDLUs per tissue area, 0.57, 95% CI [0.19, 0.77] for median TDLU span, and 0.80, 95% CI [0.62, 0.89] for median acini count per TDLU. TDLU involution measures evaluated by manual and automated assessment were inversely associated with age and menopausal status.

We have developed a computational pathology method to measure TDLU involution. This technology eliminates the labor-intensiveness and subjectivity of manual TDLU assessment, and can be applied to future breast cancer risk studies.

## Background

Most benign breast lesions and breast cancers arise in the terminal duct lobular units (TDLUs) [1], the milk-producing structures of the breast. Russo *et al.* [2] historically classified TDLUs into four lobule types: type 1 (least developed; <12 acini/lobule), type 2 (evolves from type 1; intermediate in degree of differentiation; between 12 and 80 acini/lobule), type 3 (fully developed structures; >80 acini/lobule), and type 4 (occurs during pregnancy and lactation). Pathologists have used these qualitative lobule types to evaluate TDLU involution where types 2 and 3 lobules regress back to type 1 after the completion of childbearing and with physiological aging [3]. In essence, TDLU involution is characterized by a reduction of the size of TDLUs, the number of acini, and the number of acini per TDLU [4-8]. Previous work by our group and others demonstrated that women with less TDLU involution (i.e., majority of lobules were of types 2 and 3) are more likely to develop breast cancer compared to those with predominantly type 1 lobules independent of age [5, 9, 10, 11]. Thus, TDLU involution may be utilized as a biomarker to assess breast cancer risk [9, 10].

Efforts to develop quantitative measures of TDLU involution started with McKain *et al.* [11] who evaluated the number of acini and TDLU area on histopathological sections. Rosebrock *et al.* [12] were the first to automatically estimate quantitative measurements from TDLUs and use those measurements to describe and classify them. Later, Figueroa *et al.* standardized three quantitative measures of TDLU involution—number of TDLUs per tissue area (TDLUs/mm$^2$), median TDLU span, and the median number of acini per TDLU (median acini/TDLU)—by assessing up to 10 TDLUs in the normal tissue for a WSI [4, 10, 13, 14]. These quantitative measurements still relied on manual histological assessment of breast tissue, and remained subjective and labor-intensive. Thus, the need for manual qualitative and/or quantitative assessment by pathologists is a major bottleneck to studying TDLU involution in large epidemiological studies.

Automated image analysis methods have been successfully developed for tasks in breast histopathology [15-19]. Most recently, state-of-the-art deep convolutional neural networks have been shown to outperform pathologists in detecting metastases in sentinel lymph nodes of breast cancer patients [20]. In this study, we developed an automated method to assess TDLU involution. First, we constructed and optimized three deep neural networks to detect and/or segment acini, TDLUs, and adipose tissue. These three networks were integrated into a single method to compute TDLU involution measures. Our automated method was validated by comparing the automated measures with manually acquired measures on an independent set of images.

# Methods

## Subjects and acquisition of images

The participants in this study are from the Nurses' Health Study (NHS) and NHSII. The NHS was established in 1976 with 121,700 US female registered nurses between 30-55 years of age, and NHSII was established in 1989 (n=116,429, ages 25-42). All NHS/NHSII participants are followed up biennially to obtain updated information on a range of epidemiological data and identify newly diagnosed diseases [21]. Hematoxylin and eosin (H&E) breast tissue slides were retrieved for women who reported a biopsy-confirmed benign breast disease (BBD) and gave permission to review their biopsy records and original H&E slides [22-28]. The tissue was prepared and stained at the local centers and centrally reviewed. BBD H&E whole slide images (WSIs) were obtained by scanning the slides at ×40 magnification with a resolution of $0.16\ \mu m$ per pixel using Pannoramic SCAN 150 (3DHISTECH Ltd, Budapest, Hungary). The study protocol was approved by the institutional review boards of the Brigham and Women's Hospital and Harvard T.H. Chan School of Public Health, and those of participating registries as required. Informed consent was obtained from all NHS/NHSII participants.

## Developing the automated method for TDLU involution measures

Ninety-two WSIs from 92 benign breast biopsies from 67 pre- and 25 post-menopausal women were randomly selected from the NHS database. Due to the more challenging nature of the TDLU segmentation task, 92WSIs were used to develop the TDLU segmentation neural network model while a subset of 50 out of the 92 WSIs was adequate to develop the acini detection and adipose tissue segmentation neural network models. Breast tissue with more adipose tissue has fewer

TDLUs and acini [4], which influences the outcomes of TDLU involution measures (e.g. number of TDLUs per tissue area). Therefore, the adipose tissue model was developed to estimate and account for the percentage of adipose tissue.

TDLUs, acini, and adipose tissue were annotated within a region of interest (ROI) comprising approximately 10%, 10%, and 2.5% of the total tissue area, respectively. Annotation was done using the open-source software Automated Slide Analysis Platform (ASAP; Computation Pathology Group, Radboud University Medical Center). TDLUs were defined as clusters of acini in a lobular configuration. TDLU boundary was defined by the non-specialized/extra-lobular stroma. In order to assess involution in histologically normal breast parenchyma only, TDLUs with proliferative or metaplastic changes were not annotated as TDLUs but remained as background. Acini were defined as small spherical structures lined by epithelial cells and surrounded by myoepithelial cells. Acini with elongated shapes, epithelial proliferation, apocrine metaplasia, or without lumina were not annotated. In total, 25,645 acini and 1,631 TDLUs were annotated. Figure 1 shows examples of annotated acini, TDLUs and adipose tissue.

Acini, TDLUs, and adipose tissue were detected and segmented using the U-Net convolutional neural network architecture [29, 30]. To construct the acini and adipose networks, the 50 annotated WSIs were split into training (30 WSI; 15,058 annotated acini; 60%), validation (10 WSIs; 3,561 annotated acini; 20%), and test sets (10 WSIs; 7,583 annotated acini; 20%). Annotated WSIs to construct the TDLU network were split into training (72 WSIs; 1,270 annotated TDLUs; 78%), validation (10 WSIs; 158 annotated TDLUs; 11%), and test sets (10

WSIs; 203 annotated TDLUs; 11%). The TDLU and adipose tissue segmentation models are described in Supplemental Methods. The acini detection network has been previously described [16]. To assess whether the training sets were large enough to learn to detect acini and segment TDLU ablation experiments were performed.

The three individual networks were integrated into a single automated method. This method can determine the three standardized quantitative measures by Figueroa *et al.* (i.e., TDLUs/mm$^2$, median TDLU span (µm), and median acini/TDLU [4, 10, 13, 14]) as well as two additional quantitative measures: number of acini per tissue area (acini/mm$^2$) and median TDLU area (mm$^2$). Our method can also perform TDLU involution assessment using qualitative categories as described by Russo *et al.* [2] (i.e., predominant lobule type 1, 2 or 3) and Baer *et al.* [9] (i.e., no type 1 lobules, predominantly type 1 and no type 3, and mixed lobules (all others)). Thus, in total, our automated method can capture five quantitative and two qualitative measures of TDLU involution.

**Validating the automated measures of TDLU involution**

We validated our automated method by comparing automated results with manual assessment on an independent set of 40 WSIs (Table 1). Sixty WSIs were initially chosen at random from the NHS/NHSII BBD cases to contain 30 pre- and 30 post-menopausal women. Upon further review, we excluded one woman who had type 4 lobules which suggests that she was pregnant or lactating at time of BBD diagnosis. By excluding type 4 lobules, our method is generalizable to non-pregnant/not lactating women.

For manual assessment (n=59 WSIs), two observers assessed the three standardized quantitative measures. Each observer randomly selected a ROI of approximately 50 mm$^2$ that contained an adequate number of normal TDLUs. Within the ROI, the observers estimated the percentage of breast tissue (0 to 100%) and tissue containing adipose cells (<25%, 25-50%, 50-75%, or >75%), counted the total number of TDLUs, and randomly selected up to 10 normal TDLUs to measure span (μm) and count the number of spherical acini. TDLU boundary was defined by non-specialized/extra-lobular stroma. TDLUs were not counted if >50% of their acini were dilated by 2- to 3- fold, had metaplastic changes, or displayed ductal hyperplasia. TDLUs with <50% dilated acini were included and the acini within these TDLUs were counted (including dilated ones). Acini with elongated shape or no lumen were excluded. Three observers performed qualitative assessments using predominant lobule type by Russo *et al.* [2] and categories by Baer *et al.* [9]. For intra-observer evaluation, 10 out of 40 WSIs were randomly chosen for re-assessment.

Preliminary analyses of the 59 WSIs showed that although the manual and automated TDLU assessments were highly correlated, the values of the automated results for the number of acini per TDLU were lower than manual results. Therefore, we randomly selected 19 WSIs and linear regression to derive calibration weights based on the manual results to adjust our automated results. This calibration produced more meaningful values for interpretation. We applied the calibration weights to our automated results on the remaining 40 WSIs. Tissue area was adjusted for the percentage of adipose tissue by multiplying the total tissue area by the percentage of non-adipose tissue.

**Association of TDLU measures with age and menopausal status**

We also assessed manual and automated TDLU involution measures with age and menopausal status in the final 40 cases. This was to confirm that our measures were reflective of TDLU involution, as older women were expected to have more involution.

**Statistical analysis**

The evaluation of the acini detection neural network model was done using the F1 score and the evaluation of the TDLU and adipose tissue segmentation network models was done using the Dice similarity coefficient. F1 score is the harmonic mean of precision (i.e., sensitivity) and recall (i.e., positive predictive value), which assesses how accurate the automated detection compares with ground truth (i.e., manual annotation). The calculation for Dice similarity coefficient is identical to F1 score, except it assesses the accuracy of the automated segmentation when compared to ground truth.

Inter- and intra-observer agreements for quantitative measures were summarized using intraclass correlation coefficient (ICC). Two-way mixed effects, consistency, single rater (ICC (3,1)) was used. ICC values of <0.5, between 0.5 and 0.75, between 0.75 and 0.9, and >0.9 are indicative of poor, moderate, good, and excellent reliability, respectively [31]. Intra- and inter-observer agreements for qualitative measures were determined by Fleiss' Kappa. For comparison with automated results, the consensus of the three observers was used. The consensus was determined by majority voting.

To determine the strength and direction of association of quantitative TDLU involution measures with age, Spearman's rank correlation coefficient was used. The Kruskal-Wallis test was used to examine the differences between groups of qualitative measures and age. Mann-Whitney U and Chi-squared tests were used to assess the independence of quantitative and qualitative TDLU involution assessment with menopausal status. The scores for F1, Dice, and Fleiss' Kappa range from 0 to 1, with 1 indicating perfect correlation. Analyses were performed using R and $p < 0.05$ was considered statistically significant. The ICC confidence intervals were calculated using the icc function in the irr R package.

# Results

## Performances of individual networks and establishing the automated method

The F1 score of the acini detection method was 0.73±0.09 [16]. The TDLU and adipose tissue segmentation methods obtained Dice similarity coefficients of 0.86±0.11 and 0.86±0.04, respectively. Ablation experiments showed that the methods converged with increasing number of training samples (Supplementary Figure 1).

Based on this quantitative evaluation, which indicates good agreement, and subsequent qualitative assessment we determined that the performances of these three networks were adequate to be integrated into one automated method (Figures 2 and 3; Supplementary Figure 2).

The primary cause of discordance between manual assessment and the automated method was the detection of acini and TDLU with proliferative or metaplastic changes which were intentionally excluded from manual annotation. For example, in Supplementary Figure 2C, our method incorrectly segments intraductal papillomas as TDLUs despite correctly identifying other TDLUs.

## Quantitative measures: calibration, and intra- and inter-observer agreement

The calibration coefficient to adjust the automated number of acini per TDLU measure to the manual results was found to be 3.888. The bias term was not significantly different from zero. We applied the calibration coefficient our automated results on the remaining 40 WSIs by multiplying all median number of acini per TDLU outcomes by 3.888.

Overall, quantitative measures derived from automated and manual methods achieved moderate to good inter-observer agreement (Table 2). The intra-observer agreement was good to excellent (ICC scores >0.8, 95% CI [0.53, 0.99]) and the inter-observer agreement among the two observers was moderate to good (ICC scores >0.7, 95% CI [0.51, 0.90]). The inter-observer agreement between the observers and the automated method was also moderate to good (ICC scores >0.5, 95% CI [0.19, 0.90]).

**Qualitative measures: intra- and inter-observer agreement**

Qualitative measures between the three observers and the automated method achieved fair to moderate agreement (Table 3). Among the three observers, the inter-observer Kappa scores were fair to moderate ($\kappa$ > 0.35 ($p$<0.01)) while there was a large variation in their intra-observer Kappa scores ($\kappa$ from 0.048 ($p$=0.880) to 1.000 ($p$<0.01)). The inter-observer agreement between the observers and the automated method was moderate ($\kappa$ > 0.5 ($p$<0.01)). There was slightly more agreement in the evaluation of Russo *et al.* [2] predominant lobule type compared to Baer *et al.* [9] categories.

**TDLU involution with age and menopausal status**

All quantitative and qualitative measures obtained by manual and automated methods were significantly associated with age ($p$<0.05; Figures 4 and 5). Table 4 shows the relationships between TDLU measures and menopausal status. All quantitative measures were significantly different between pre- and post-menopausal women, except number of TDLUs per tissue area evaluated by the automated method ($p$=0.06). Likewise, qualitative measures (consensus vote by observers and automated method) were significantly different between pre- and post-menopausal women, except lobular classification according to Baer *et al.* [2] assessed by the automated method ($p$=0.07). No participant was classified as predominantly type 3 according to Russo *et al.* [2]. Qualitative measures when assessed by individual observers were not associated with menopausal status ($p$>0.05; Supplementary Table 1).

Thus, older and post-menopausal women had significantly fewer TDLUs/mm$^2$, smaller TDLUs, reduced number of acini per TDLU, and fewer acini/mm$^2$ compared to pre-menopausal women. Type 1 lobules were predominantly observed in post-menopausal women while the majority of pre-menopausal women had mixed lobules.

## Discussion

Greater amounts of TDLU involution are inversely associated with breast cancer risk [5, 6, 9-11] and aggressive breast cancer subtypes [13, 14]. It is important to better understand TDLU involution as well as epidemiological factors that influence the involution process to obtain deeper insights into breast carcinogenesis and identify new opportunities for breast cancer prevention. A major bottleneck to studying TDLU involution and breast cancer risk in large epidemiological cohorts is the need for manual qualitative and/or quantitative assessment by pathologists. In this study, we developed and validated a computational pathology method that can assess five quantitative and two qualitative measures of TDLU involution. Our automated method was highly comparable to manual assessment, and we confirmed that our TDLU involution measures reflect age and menopausal status [4]. This technology will be a valuable research tool to facilitate future breast cancer risk studies.

Our automated method integrates three separate networks for acini detection, TDLU segmentation, and adipose tissue segmentation. It was challenging to develop the TDLU segmentation network compared to the other two networks because TDLUs have highly variable appearances and BBD encompasses a wide range of morphology. As such, the TDLU segmentation network required more training WSIs to achieve a Dice score similar to the adipose tissue segmentation network. Since we are the first to develop networks for acini detection and TDLU segmentation, we were unable to benchmark our networks. We identified three primary causes of discordance between manual assessment and the automated method which affected our F1 and Dice scores: 1) acini with proliferative or metaplastic changes were frequently detected by the network but were intentionally excluded from manual annotation; 2) the network had difficulty predicting boundaries of TDLUs with complex clustering; and 3) in some cases, the network interpreted large ducts as adipose tissue.

Despite researchers' best efforts to create a perfect method, most automated methods remain prone to segmentation errors. Solutions to address these issues and improve our computational method include increasing the number of training samples with improved annotation and applying hard negative mining. The inclusion of abnormal epithelium when assessing TDLU involution may influence breast cancer risk assessment. Therefore, future work will evaluate the inter-variability of TDLU measures between slides obtained from different tissue blocks for each patient. In addition, summarizing the automated results using median instead of mean, and evaluating at least two WSIs per case (averaging the median values), will improve the robustness and reliability of the data in future studies.

To capture TDLU span, the automated method uses the length of the major axis of the ellipse that is identical to the normalized second central moments for each TDLU. In contrast, a pathologist has to select two opposite points along the boundary of a TDLU to obtain the longest span. Thus, the manual assessment of TDLU span inevitably contains some subjectivity and explains the low inter-observer agreement score between manual and automated results. Our automated method has the ability to capture two new measures: number of acini per tissue area and median TDLU area. Future studies will evaluate and compare these newer measures with the existing three standardized measures to determine which TDLU involution quantitative measure is most associated with breast cancer risk.

TDLU involution is historically assessed using qualitative measures [2, 5, 9]. The large variation in intra- and inter-observer Kappa scores as observed in this study reiterated the high subjectivity of qualitative measures, thus spurring researchers to develop standardized quantitative measures to assess TDLU involution [4, 10, 13, 14]. Our study showed higher intra- and inter-observer agreement for quantitative measures compared to qualitative measures, again highlighting the reproducibility of quantitative measures. Despite assessing different tissue areas for manual assessment (observers selected 50 $mm^2$ tissue area) and automated method (entire tissue area on WSI), the good agreement between the observers and automated results provided additional assurance that our automated method is comparable to manual assessment.

# Conclusion

We developed and validated an automated method to measure TDLU involution as a first step towards automated prediction of breast cancer risk. Qualitative assessment of TDLU involution is a subjective process. Quantitative assessment produces more reproducible results but is labor-intensive for pathologists. Our method can eliminate the labor-intensiveness and subjectivity of manual TDLU involution assessment. Our technology can be applied on a larger scale to assess breast cancer risk in epidemiological studies. Future work will determine the best quantitative TDLU involution measure to predict breast cancer risk, and evaluate the impact of incorporating these measures into clinical breast cancer risk assessment models to improve patient management.

**Competing interests**

The authors declare no competing interests.

**Author's contributions**

Conceived and designed the study: YJH MV SCW RMT GMB LCC SJS. Collection of NHS data: RMT KHK YL YJH MEP. Pathological annotation and TDLU involution assessment: AMO CL GMB LCC. Image processing, development and implementation of the automated method: SCW MV JPWP MEP BB RV RVL. Data analyses: SCW MV YJH. All authors contributed to the writing and reviewing of the manuscript.

**Acknowledgements**

We would like to thank the participants and staff of the Nurses' Health Study and Nurses' Health Study II for their valuable contributions as well as the following state cancer registries for their help: AL, AZ, AR, CA, CO, CT, DE, FL, GA, ID, IL, IN, IA, KY, LA, ME, MD, MA, MI, NE, NH, NJ, NY, NC, ND, OH, OK, OR, PA, RI, SC, TN, TX, VA, WA, WY. The authors assume full responsibility for analyses and interpretation of these data. This work was supported by the National Institute of Health/National Cancer Institute R21CA187642 (RMT), UM1CA186107, and UM1CA176726, the Susan G. Komen Foundation (RMT), the Klarman Family Foundation (YJH), and the Deep Learning for Medical Image Analysis research program by Netherlands Organization for Scientific Research and Philips Research P15-26 (SCW, MV and JPWP).

**Data availability statement**

Subject clinical data, whole slide images, and pathological reviews, that support the findings of this study are not publicly available. Data are available from the Nurses' Health Studies. Investigators interested in using the data can request access, and feasibility will be discussed at an investigators meeting. Limits are not placed on scientific questions or methods, and there is no requirement for co-authorship. Additional data sharing information and policy details can be accessed at http://www.nurseshealthstudy.org/researchers.

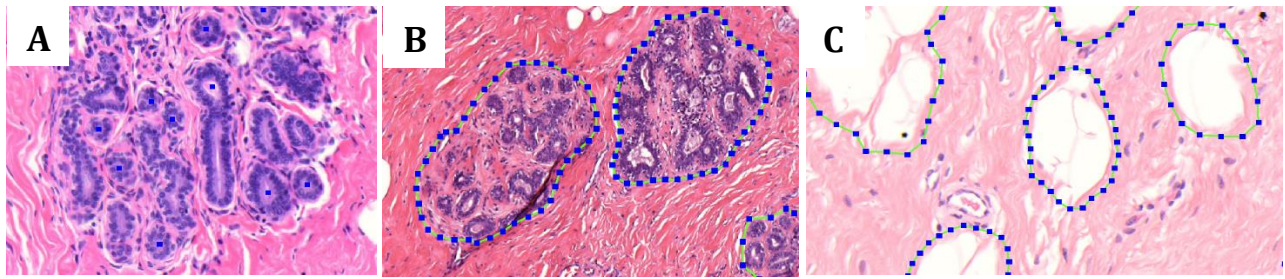

**Figure 1:** Examples of annotations for acini (**A;** annotated by blue squares), terminal duct lobular units (**B**) and adipose tissue (**C**).

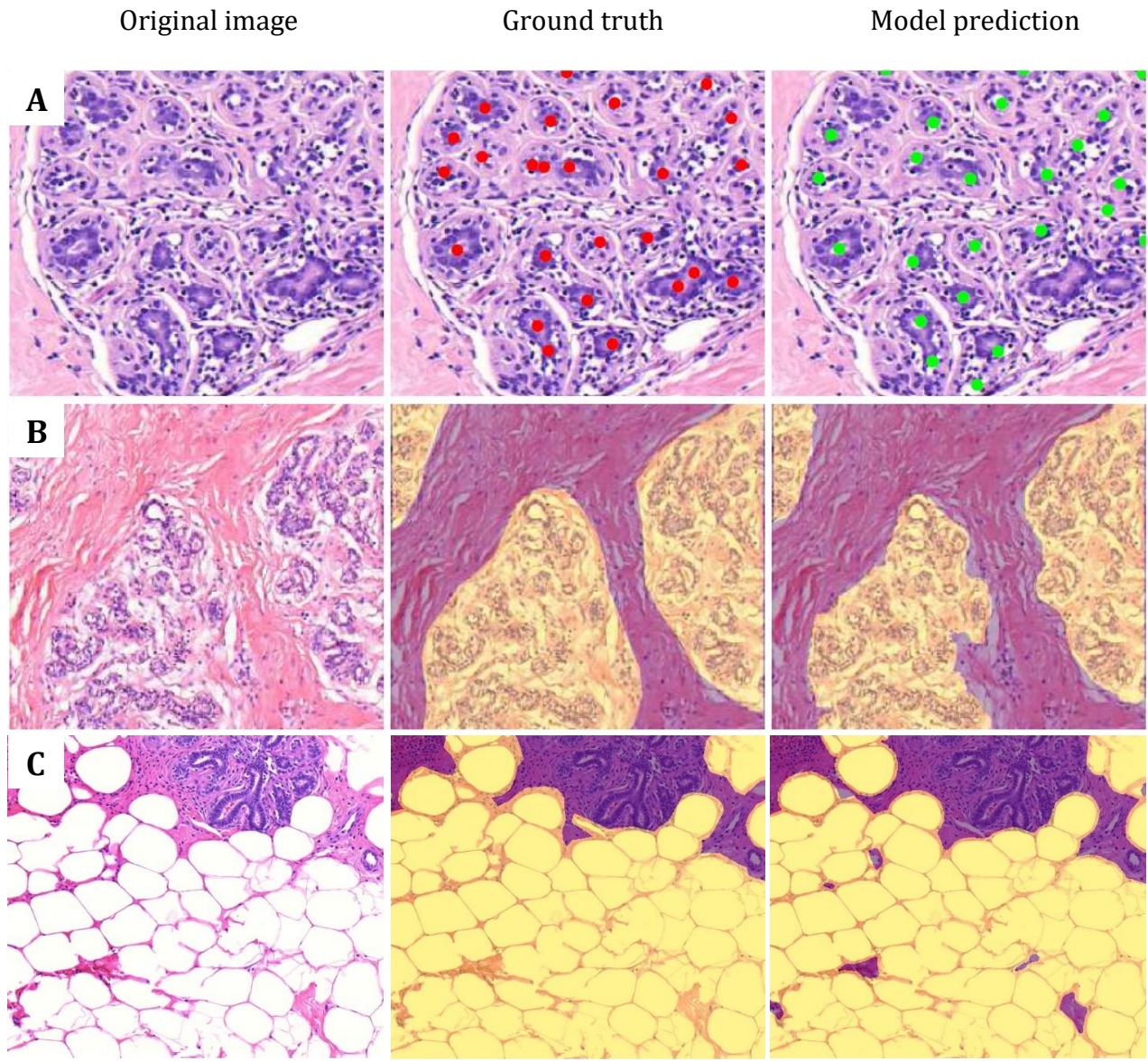

**Figure 2:** Results of the acini detection (**A**), terminal duct lobular unit (**B**), and adipose tissue (**C**) segmentation algorithms. The original images are in the left column, the middle column shows ground truth as annotated by human observers, and the detections and segmentations performed by the automated method are displayed in the right column.

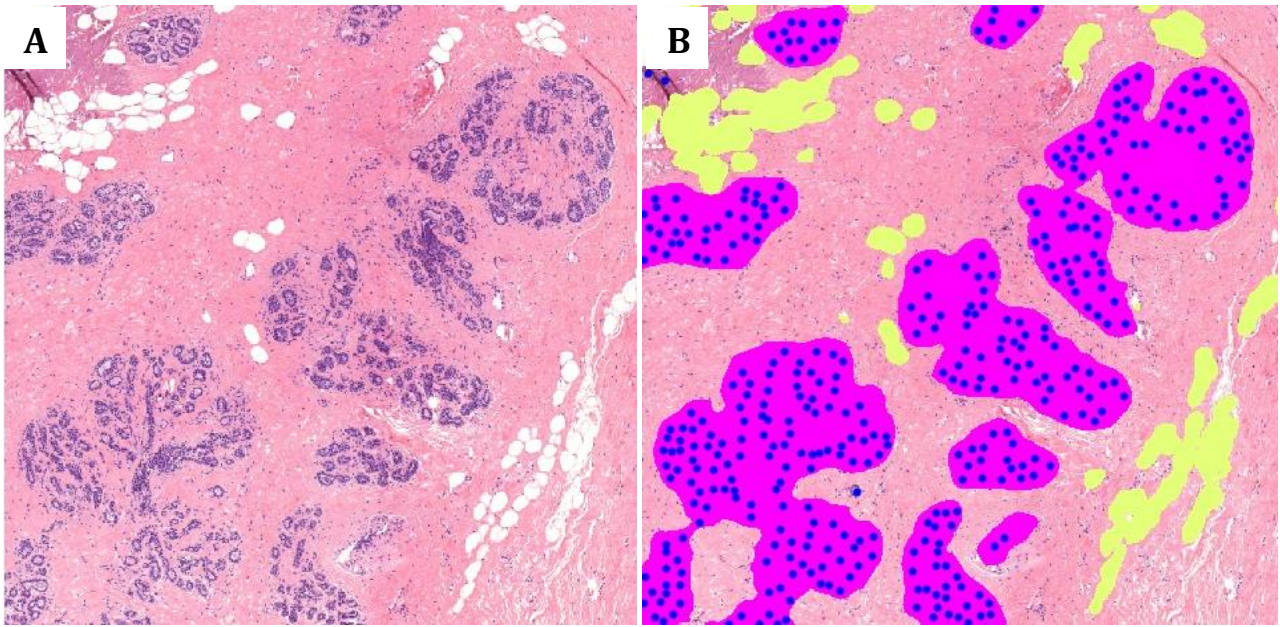

**Figure 3:** Results of the acini detection, terminal duct lobular unit, and adipose tissue segmentation algorithms (**B**) overlaid on the original image (**A**).

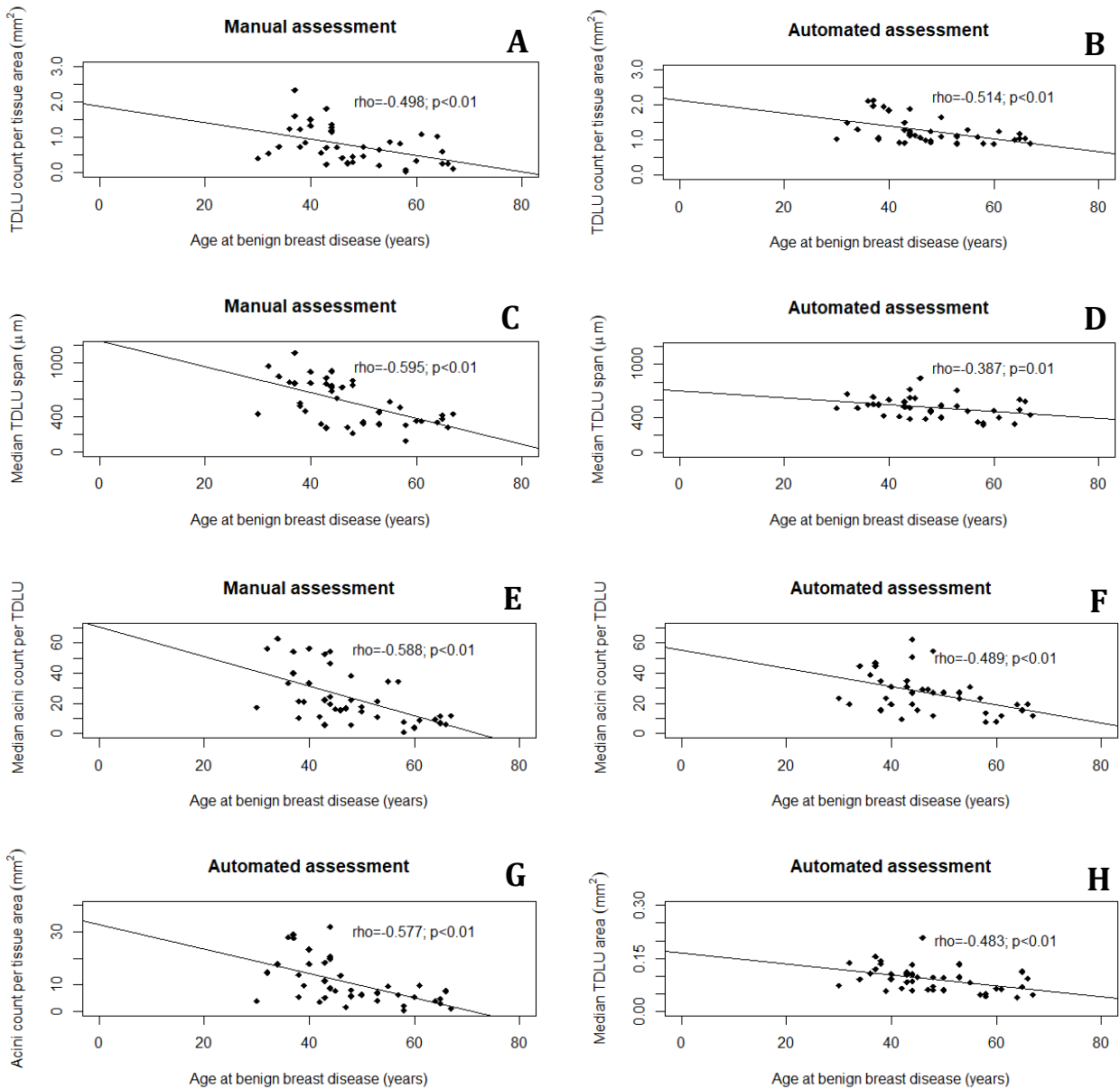

**Figure 4:** Scatterplots of the association of quantitative terminal ductal lobular unit (TDLU) involution measures and age. TDLU count per tissue area assessed using manual (**A**) and automated (**B**) method were significantly inversely correlated with age (*p*<0.01). Median TDLU span assessed manually (**C**) and with the automated method (**D**) was significantly inversely correlated with age (*p*<0.01 and *p*=0.01). Median acini count per TDLU assessed using manual (**E**) and automated (**F**) assessment was also significantly inversely correlated with age (*p*<0.01). Acini count per tissue area assessed by the automated method was significantly inversely correlated with age (**G**; *p*<0.01). Median TDLU area assessed by the automated method was significantly inversely correlated with age (**H**; *p*<0.01).

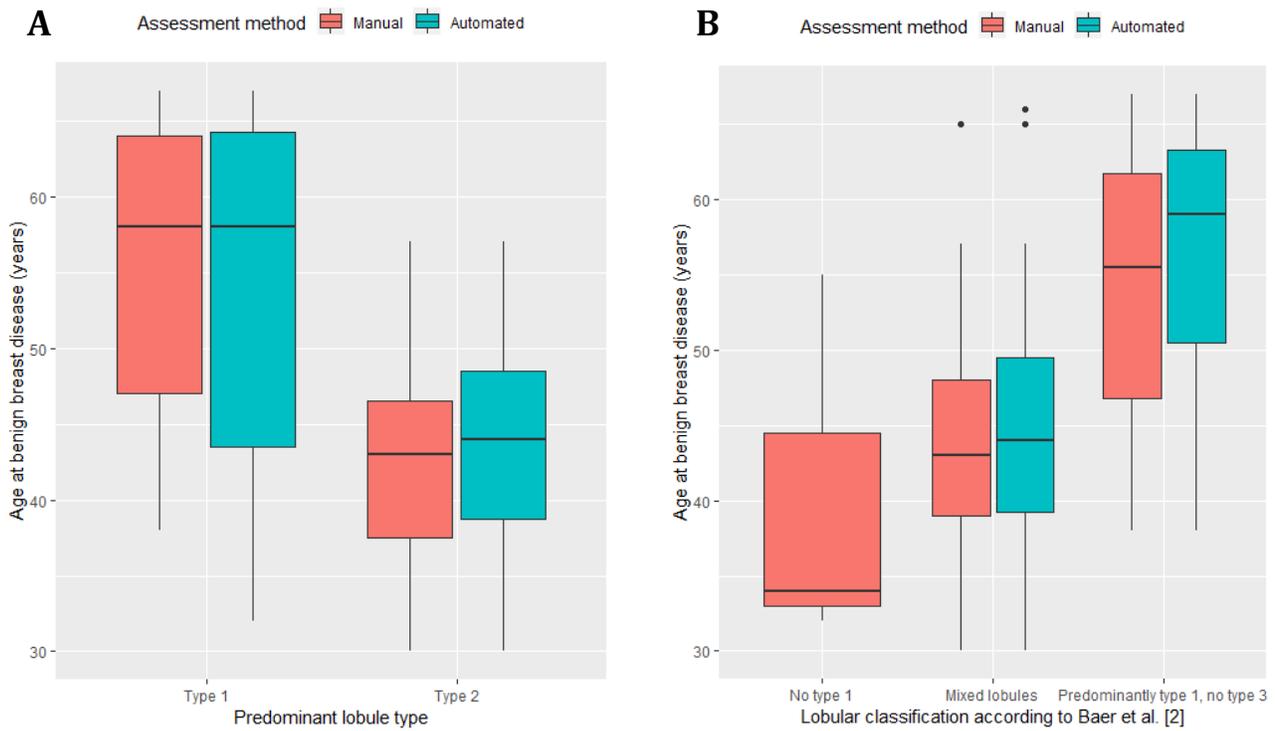

**Figure 5:** Boxplots demonstrating the association of qualitative terminal ductal lobular unit involution measures and age. (**A**) Women with predominantly type 1 lobules were significantly older than women with predominantly type 2 lobules (manual method: $p<0.01$; automated method: $p=0.01$). No woman presented with predominately type 3 lobules. (**B**) Women with "Predominantly type 1, no type 3" lobules were significantly older than women with "Mixed lobules" (manual method $p<0.01$; automated method $p<0.01$). No woman was assessed as having "No type 1" lobules by the automated method. The manual qualitative measures were obtained by consensus vote. The boxplots show the median value, interquartile range (IQR), and 5th and 95th whiskers.

**Table 1:** Demographic table of 40 participants used to validate the automated measures of TDLU involution.

|  | Pre-Menopausal | Post-Menopausal |
|---|---|---|
| *n* | 20 | 20 |
| Cohort, *n (%)* | | |
| Nurses' Health Study | 5 (25) | 12 (60) |
| Nurses' Health Study II | 15 (75) | 8 (40) |
| Year of benign breast disease diagnosis, *n (%)* | | |
| ≥1978 to <1988 | 3 (15) | 4 (20) |
| ≥1988 to <1998 | 16 (80) | 12 (60) |
| ≥1998 to 2000 | 1 (5) | 4 (20) |
| Age at benign breast disease diagnosis, *n (%)* | | |
| 30 to 39 | 8 (40) | 1 (5) |
| 40 to 49 | 10 (50) | 6 (30) |
| 50 to 59 | 2 (10) | 6 (30) |
| ≥60 | 0 (0) | 7 (35) |

**Table 2:** Inter- and intra-observer intraclass correlation coefficient (ICC) scores and the 95% confidence interval (CI) for the quantitative terminal ductal lobular unit involution measures obtained from two observers and the automated method.

| | Intra-observer ICC (95% CI)* | | Inter-observer ICC (95% CI)# | |
| --- | --- | --- | --- | --- |
| | Observer 1 | Observer 2 | Observer 1 vs 2 | mean(observers) vs automated |
| Number of TDLUs per tissue area (mm$^2$) | 0.96 (0.86, 0.99) | 0.82 (0.78, 0.98) | 0.71 (0.51, 0.83) | 0.80 (0.63, 0.90) |
| Median TDLU span ($\mu$m) | 0.91 (0.69, 0.98) | 0.90 (0.67, 0.98) | 0.81 (0.67, 0.90) | 0.57 (0.19, 0.77) |
| Median number of acini per TDLU | 0.91 (0.69, 0.98) | 0.86 (0.53, 0.96) | 0.73 (0.54, 0.85) | 0.80 (0.62, 0.89) |

*Intra-observer ICC was evaluated using 10 out of the 40 cases.
#Inter-observer ICC was evaluated using 40 cases.

**Table 3:** Inter- and intra-observer Fleiss' Kappa for qualitative terminal ductal lobular unit assessment among three observers using 40 and 10 cases, respectively.

| | Intra-observer* | | | | | | Inter-observer# | | | |
|---|---|---|---|---|---|---|---|---|---|---|
| | Observer 1 | | Observer 2 | | Observer 3 | | Observer 1,2 & 3 | | Consensus vote of observers vs automated | |
| | $\kappa$ | p-value | $\kappa$ | p-value | $\kappa$ | p-value | $\kappa$ | p-value | $\kappa$ | p-value |
| Predominant lobular type by Russo *et al.* [2] | 0.167 | 0.598 | 0.608 | 0.055 | 0.798 | **0.012** | 0.529 | **<0.01** | 0.536 | **<0.01** |
| Lobular classification according to Baer *et al.* [9] | 0.048 | 0.880 | 1.000 | **<0.01** | 0.798 | **0.012** | 0.370 | **<0.01** | 0.538 | **<0.01** |

*Intra-observer evaluation was done using 10 out of the 40 cases.
#Inter-observer evaluation was done using 40 cases.

**Table 4:** The association of terminal ductal lobular unit (TDLU) involution measures and menopausal status.

| | Pre-Menopausal | Post-Menopausal | p-value |
|---|---|---|---|
| *n* | 20 | 20 | |
| **Quantitative measures** | | | |
| Number of TDLU per tissue area (mm²), median *n (IQR)* | | | |
|   Evaluated by observers | 0.74 (0.46,1.34) | 0.65 (0.27,0.86) | **0.04** |
|   Evaluated by the automated method | 1.19 (1.05,1.84) | 1.07 (0.92,1.26) | 0.06 |
| Median TDLU span in *μ*m, median *n (IQR)* | | | |
|   Evaluated by observers | 740.40 (502.35,810.02) | 362.90 (317.01,519.75) | **<0.01** |
|   Evaluated by the automated method | 536.64 (504.17,580.56) | 448.35 (392.73,587.87) | **<0.05** |
| Number of acini per TDLU, median *n (IQR)* | | | |
|   Evaluated by observers | 29.00 (16.81,48.00) | 11.75 (8.50,20.06) | **<0.01** |
|   Evaluated by the automated method | 30.13 (26.24,40.34) | 19.44 (13.12,24.30) | **<0.01** |
| Number of acini per tissue area (mm²), median *n (IQR)* | | | |
|   Evaluated by the automated method | 14.18 (6.30,20.09) | 5.75 (3.43,8.90) | **<0.01** |
| Median TDLU area (mm²), median *n (IQR)* | | | |
|   Evaluated by the automated method | 0.10 (0.08,0.12) | 0.06 (0.06,0.10) | **<0.01** |
| **Qualitative assessment** | | | |
| Predominant lobular type by observers (consensus vote), *n (%)* | | | **0.01** |
|   Type 1 | 4 (20.0) | 13 (65.0) | |
|   Type 2 | 16 (80.0) | 7 (35.0) | |
|   Type 3 | 0 (0.0) | 0 (0.0) | |
| Predominant lobular type by the automated method, *n (%)* | | | **0.02** |
|   Type 1 | 4 (20.0) | 12 (60.0) | |
|   Type 2 | 16 (80.0) | 8 (40.0) | |
|   Type 3 | 0 (0.0) | 0 (0.0) | |
| Lobular classification according to Baer *et al.* [2] by observers (consensus vote), *n (%)* | | | **0.04** |
|   No type 1 | 2 (10.0) | 1 (5.0) | |
|   Mixed lobules | 14 (70.0) | 7 (35.0) | |
|   Predominantly type 1, no type 3 | 4 (20.0) | 12 (60.0) | |
| Lobular classification according to Baer *et al.* [2] by the automated method, *n (%)* | | | 0.07 |
|   No type 1 | 0 (0.0) | 0 (0.0) | |
|   Mixed lobules | 18 (90.0) | 12 (60.0) | |
|   Predominantly type 1, no type 3 | 2 (10.0) | 8 (4 0.0) | |

# Supplementary Methods

## Convolutional neural network architecture

The neural network used to train our acini detection, TDLU segmentation and adipose tissue segmentation networks is similar to the one described by Ronneberger *et al.* [1]. We use U-Nets with the same depth and amount of filters. The U-Net architecture consists of a contracting path (as is usual in neural networks) succeeded by a symmetric expanding path. This architecture was designed to capture context yet also enable precise localization of objects.

During training patches of 512×512 pixels were extracted from the annotated part of the WSI and randomly translated and rotated. The mini batch size was set to 10 patches and the network was trained by minimizing the binary cross-entropy between the ground truth and predictions with an Adam optimizer with a learning rate of 1e-6. Training was stopped when the average of the validation loss over 10 epochs increased. Hyper parameters like the depth of the U-Net, the amount of filters, the mini batch size and the learning rate were tuned to optimal performance on the validation set by grid search. This was done for each network individually but the same parameters were optimal for all three.

## Acini detection

The acini detection network was trained and evaluated using 50 WSIs from 50 NHS/NHSII participants. Thirty WSIs were used for training, 10 for validation and 10 for testing. In these WSIs a region comprising 10% of the total tissue area was annotated. The annotations were centroid only, meaning that the center pixel was annotated and not the extent of the acinus. Since only the center pixel of the acini was annotated, taking the trivial segmentation approach would lead to a severe class imbalance. To address this, we defined alternative targets to train our deep learning architecture on. A comparison of different targets can be found in our previous work [2]. The best performing method was to use soft centroid labels, in which we place an isotropic Gaussian with a standard deviation of 10 pixels at the location of each acinus centroid.

After training, the predicted target maps were converted to acini centroid predictions by using non-maximum suppression in a radius of 20 pixels, with a threshold of 0.48 (out of 1). These hyper-parameters were determined based on the validation set. More information on this method can be found in [2].

## TDLU segmentation

The TDLU segmentation network was trained and evaluated using 92 WSIs from 92 NHS/NHSII participants. Seventy-two WSIs were used for training, 10 were used for validation and 10 for testing. In these WSIs a region comprising 10% of the total tissue area was annotated with TDLU segmentations. TDLUs were defined as clusters of acini in a lobular configuration. TDLU boundary was defined by the non-specialized/extra-lobular stroma. In order to assess involution in histologically normal breast parenchyma only, TDLUs with proliferative or metaplastic changes were not annotated.

After training, the predicted target maps were converted to TDLU segmentations by using thresholding and morphological operations. More specifically, we first calculated a threshold for the prediction map using Otsu's method [3]. All values below this threshold were set to 0. Then, we removed all connected objects in the image that had an area smaller than 2500 pixels. After that, a median filter with a kernel size of 11 was used to remove noise. Lastly, holes in the remaining objects were removed if they were smaller than 2500 pixels. These thresholding and morphological operations and their parameters were determined based on performance on the validation set.

**Adipose tissue segmentation**

The adipose tissue segmentation network was trained and evaluated using 50 WSIs from 50 NHS/NHSII participants. Fifty WSIs were used for training, 10 were used for validation and 10 for testing. In these WSIs a region comprising 2.5% of the total tissue area was annotated with adipose tissue segmentations.

After training, the predicted target maps were converted to adipose tissue segmentations by thresholding with a value of 0.6. This threshold was determined based on performance on the validation set.

# Supplementary Figures

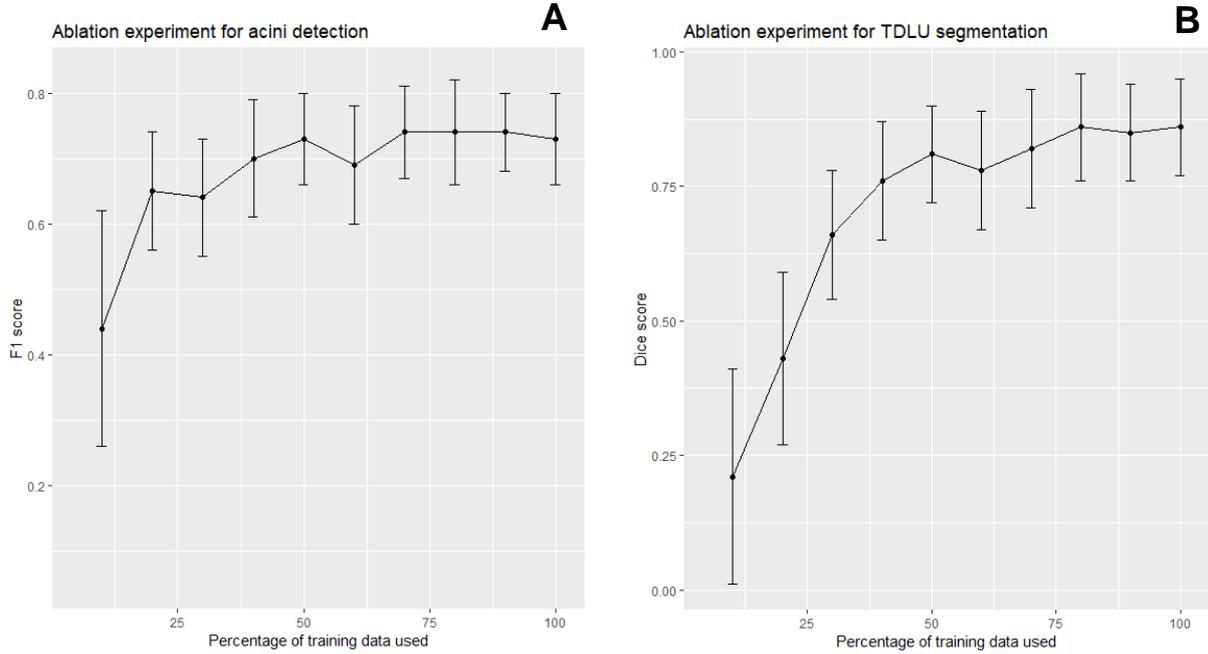

**Supplementary Figure 1:** Line charts demonstrating the F1 score obtained on the test set with models trained using different percentages of the training dataset. (**A**) The ablation experiment for the detection of acini. The line converges before it reaches 100% of the training data indicating that the training set is large enough. (**B**) The ablation experiment for the segmentation of TDLUs. The line converges before it reaches 100% of the training data indicating that the training set is large enough. The line charts show the mean value and standard deviation.

**A.1**

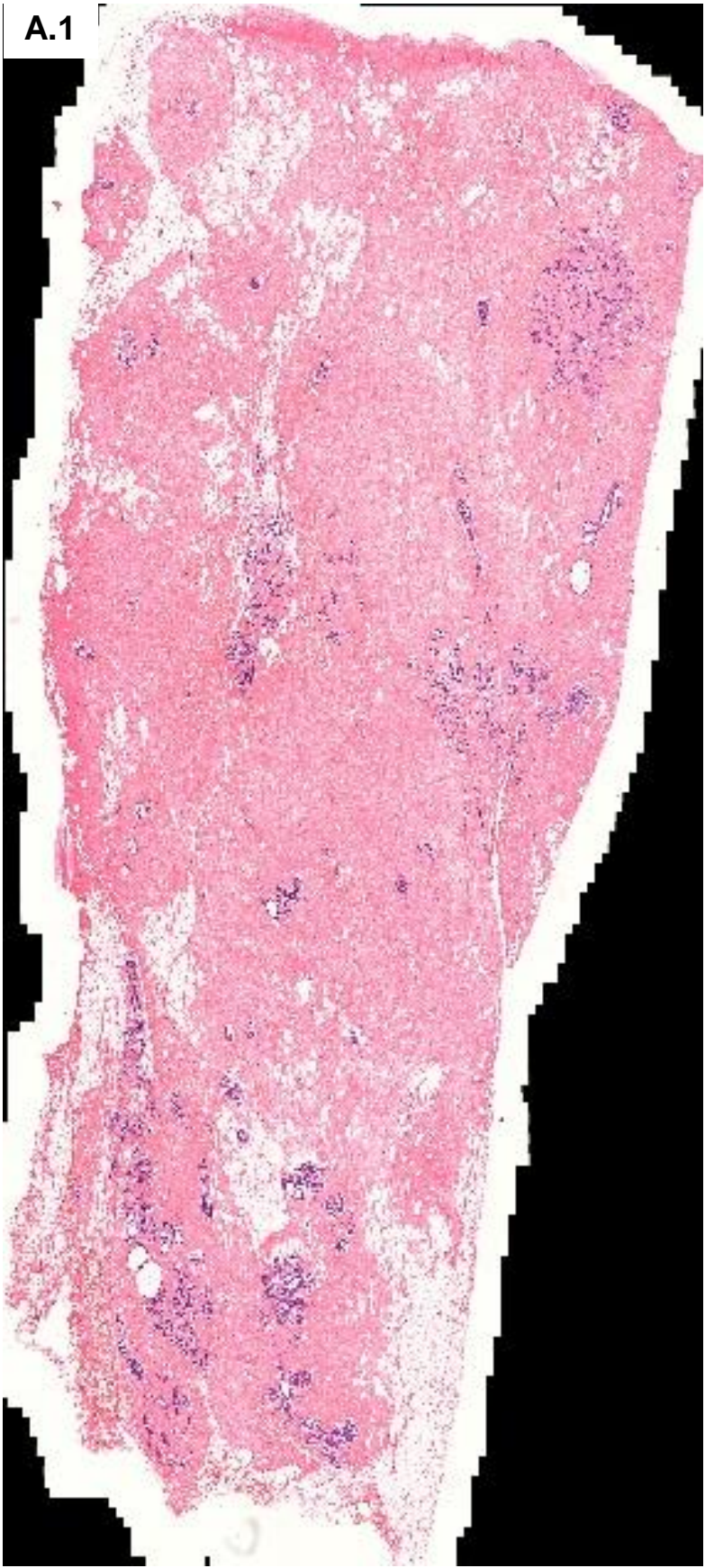

**A.2**

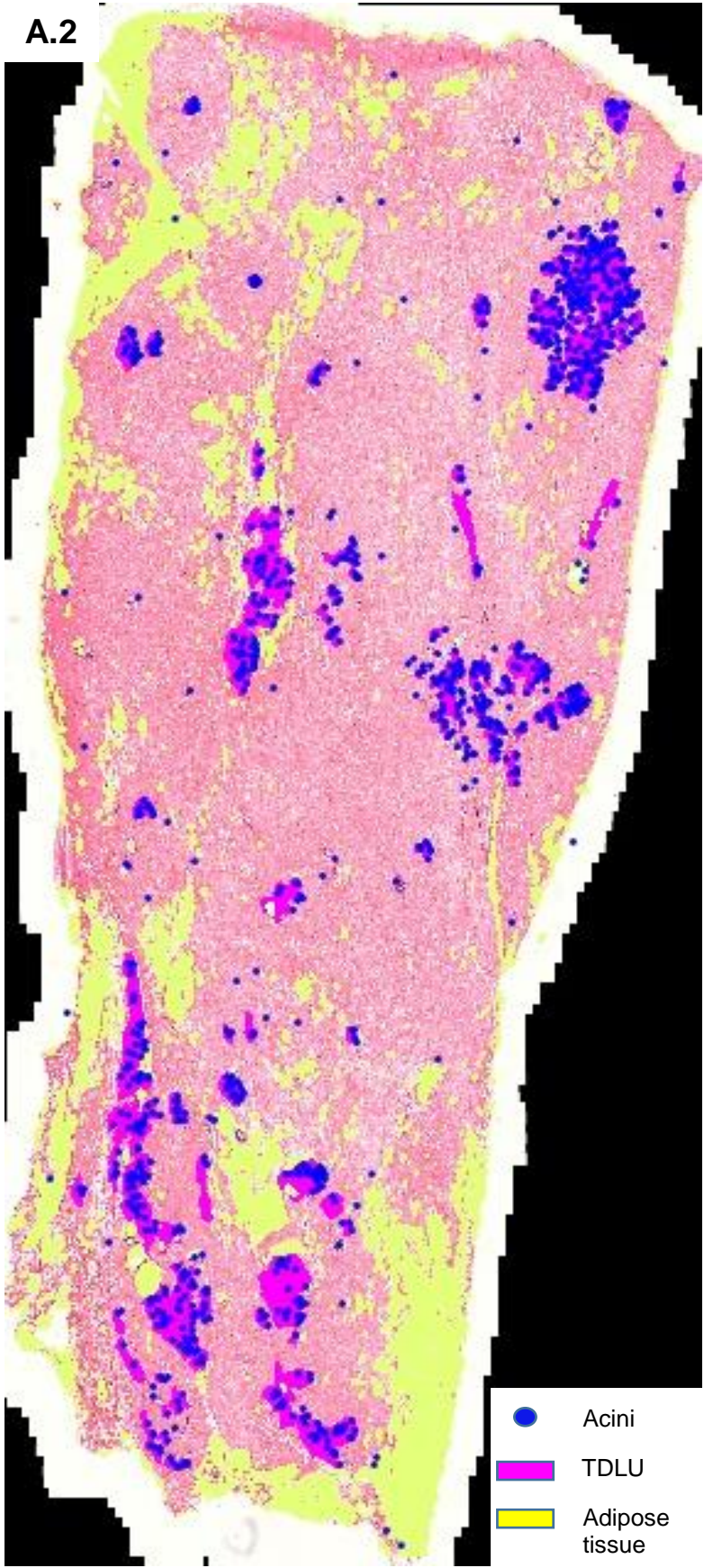

● Acini

TDLU

Adipose
tissue

**B.1**

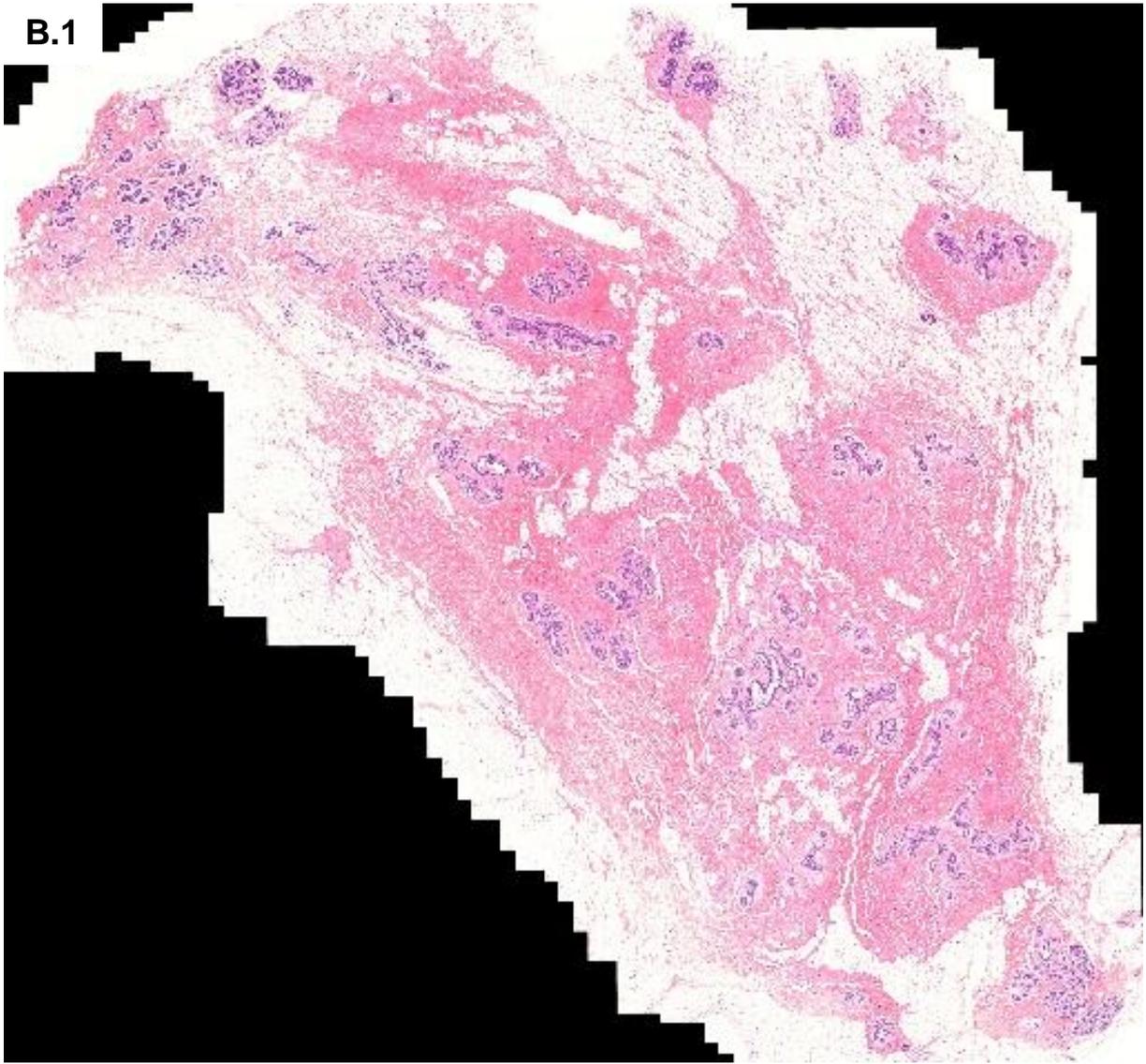

**B.2**

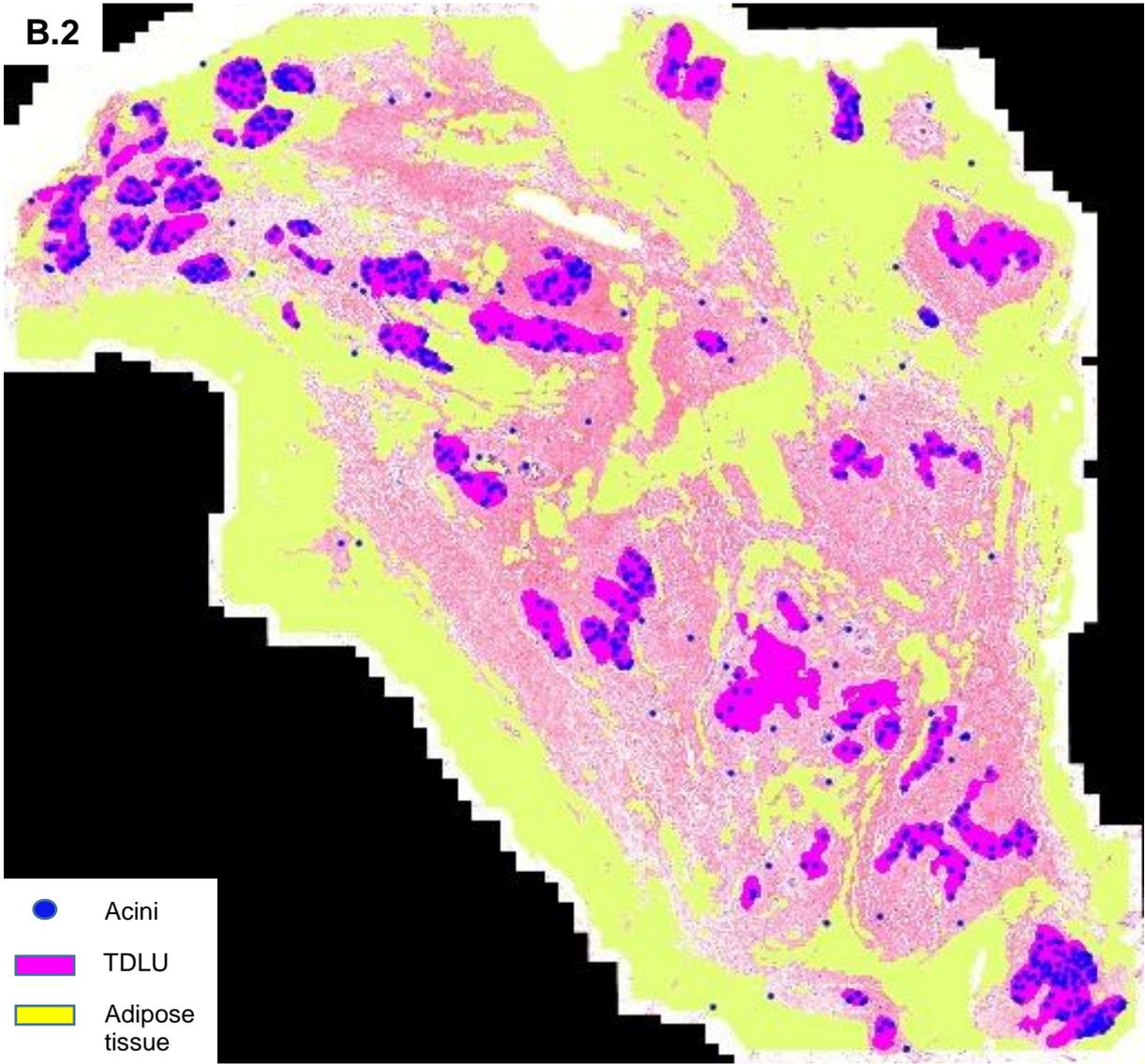

- ● Acini
- ■ TDLU
- ▢ Adipose tissue

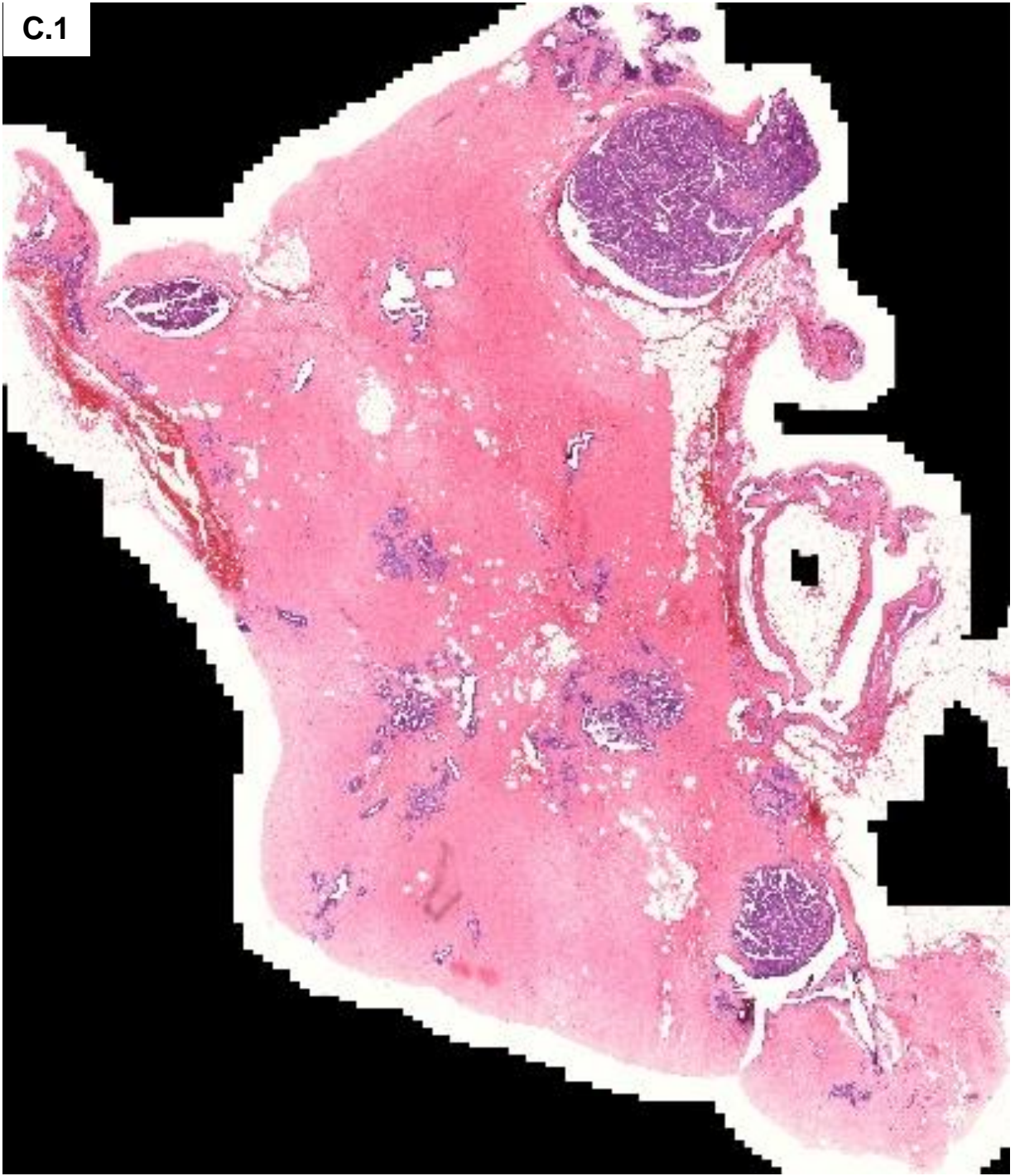

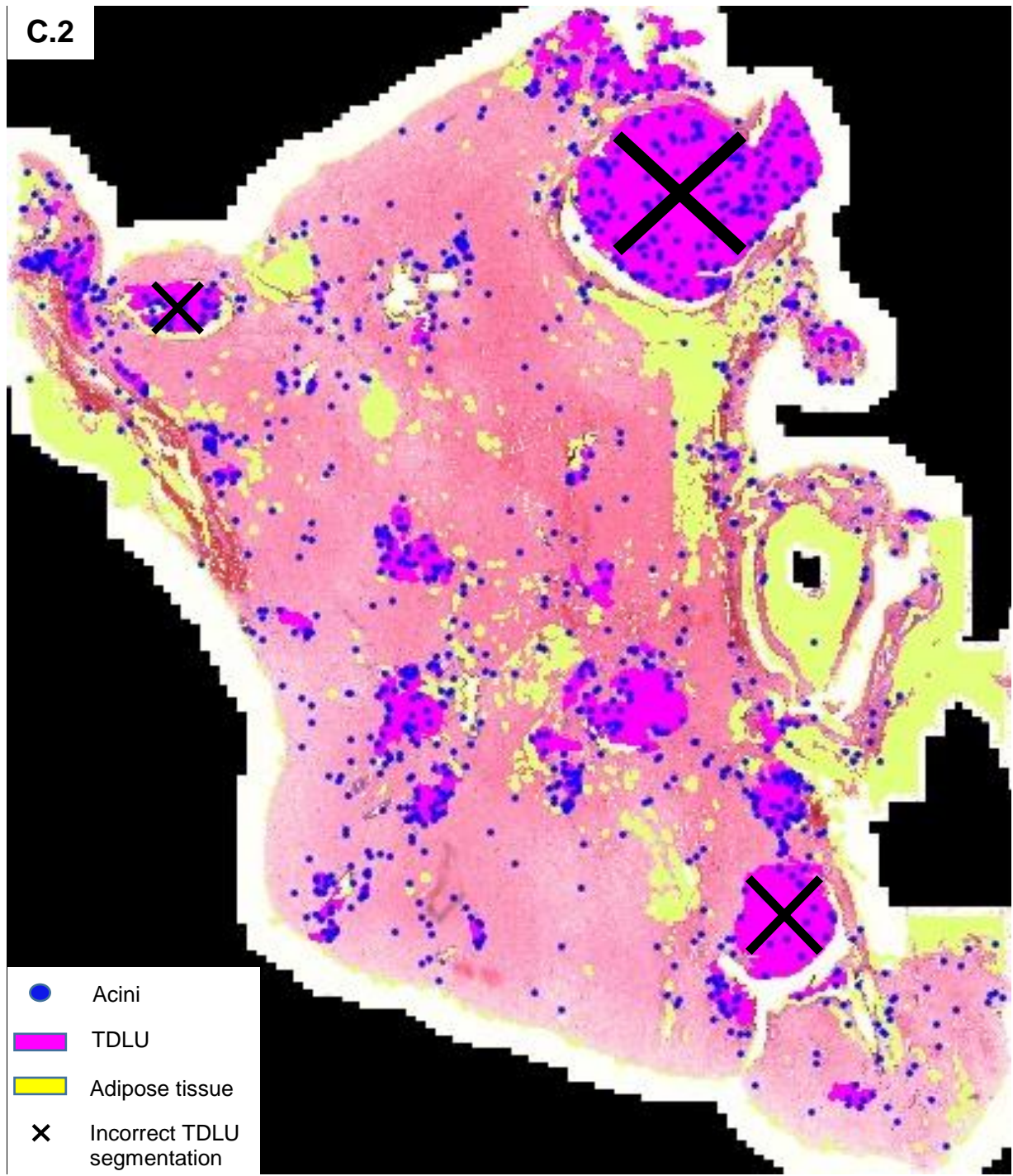

**Supplementary Figure 2:** Results of the automated method (**A.2, B.2, C.2**) overlaid on original whole slide images (**A.1, B.1, C.1**). Detected acini are shown in blue, terminal duct lobular units (TDLUs) in yellow, and adipose tissue in pink. The black crosses (**C.2**) indicate regions where intraductal papillomas were incorrectly segmented as TDLUs.

# Supplementary Tables

**Supplementary Table 1:** The association of terminal ductal lobular unit (TDLU) involution measures and menopausal status.

| | Pre-Menopausal | Post-Menopausal | p-value |
|---|---|---|---|
| *n* | 20 | 20 | |
| **Quantitative measures** | | | |
| Number of TDLU per tissue area (mm²), median *n (IQR)* | | | |
|   Evaluated by observers | 0.74 (0.46,1.34) | 0.65 (0.27,0.86) | **0.04** |
|   Evaluated by the automated method | 1.19 (1.05,1.84) | 1.07 (0.92,1.26) | 0.06 |
| Median TDLU span in $\mu$m, median *n (IQR)* | | | |
|   Evaluated by observers | 740.40 (502.35,810.02) | 362.90 (317.01,519.75) | **<0.01** |
|   Evaluated by the automated method | 536.64 (504.17,580.56) | 448.35 (392.73,587.87) | **<0.05** |
| Number of acini per TDLU, median *n (IQR)* | | | |
|   Evaluated by observers | 29.00 (16.81,48.00) | 11.75 (8.50,20.06) | **<0.01** |
|   Evaluated by the automated method | 30.13 (26.24,40.34) | 19.44 (13.12,24.30) | **<0.01** |
| Number of acini per tissue area (mm²), median *n (IQR)* | | | |
|   Evaluated by the automated method | 14.18 (6.30,20.09) | 5.75 (3.43,8.90) | **<0.01** |
| Median TDLU area (mm²), median *n (IQR)* | | | |
|   Evaluated by the automated method | 0.10 (0.08,0.12) | 0.06 (0.06,0.10) | **<0.01** |
| **Qualitative assessment** | | | |
| Predominant lobular type by observer 1, *n (%)* | | | 0.13 |
|   Type 1 | 6 (30.0) | 12 (60.0) | |
|   Type 2 | 13 (65.0) | 6 (30.0) | |
|   Type 3 | 1 (5.0) | 1 (5.0) | |
|   Missing | 0 (0.0) | 1 (5.0) | |
| Predominant lobular type by observer 2, *n (%)* | | | 0.05 |
|   Type 1 | 4 (20.0) | 11 (55.0) | |
|   Type 2 | 16 (80.0) | 9 (45.0) | |
|   Type 3 | 0 (0.0) | 0 (0.0) | |
| Predominant lobular type by observer 3, *n (%)* | | | 0.07 |
|   Type 1 | 6 (30.0) | 12 (60.0) | |
|   Type 2 | 14 (70.0) | 7 (35.0) | |
|   Type 3 | 0 (0.0) | 0 (0.0) | |
|   Missing | 0 (0.0) | 1 (5.0) | |
| Predominant lobular type by observers (consensus vote), *n (%)* | | | **0.01** |
|   Type 1 | 4 (20.0) | 13 (65.0) | |
|   Type 2 | 16 (80.0) | 7 (35.0) | |
|   Type 3 | 0 (0.0) | 0 (0.0) | |
| Predominant lobular type by the automated method, *n (%)* | | | **0.02** |
|   Type 1 | 4 (20.0) | 12 (60.0) | |
|   Type 2 | 16 (80.0) | 8 (40.0) | |
|   Type 3 | 0 (0.0) | 0 (0.0) | |
| Lobular classification according to Baer *et al.* [2] by observer 1, *n (%)* | | | 0.24 |
|   No Type 1 | 2 (10.0) | 1 (5.0) | |
|   Mixed lobules | 13 (65.0) | 8 (40.0) | |
|   Predominantly Type 1, no Type 3 | 5 (25.0) | 9 (45.0) | |
|   Missing data | 0 (0.0) | 1 (5.0) | |
| Lobular classification according to Baer *et al.* [2] by observer 2, *n (%)* | | | **<0.01** |
|   No Type 1 | 11 (55.0) | 1 (5.0) | |
|   Mixed lobules | 5 (25.0) | 9 (45.0) | |
|   Predominantly Type 1, no Type 3 | 4 (20.0) | 10 (50.0) | |
| Lobular classification according to Baer *et al.* [2] by observer 3, *n (%)* | | | **0.03** |
|   No Type 1 | 0 (0.0) | 0 (0.0) | |
|   Mixed lobules | 15 (75.0) | 7 (35.0) | |
|   Predominantly Type 1, no Type 3 | 5 (25.0) | 12 (60.0) | |
|   Missing data | 0 (0.0) | 1 (5.0) | |
| Lobular classification according to Baer *et al.* [2] by observers (consensus vote), *n (%)* | | | **0.04** |
|   No Type 1 | 2 (10.0) | 1 (5.0) | |
|   Mixed lobules | 14 (70.0) | 7 (35.0) | |
|   Predominantly Type 1, no Type 3 | 4 (20.0) | 12 (60.0) | |
| Lobular classification according to Baer *et al.* [2] by the automated method, *n (%)* | | | 0.07 |
|   No Type 1 | 0 (0.0) | 0 (0.0) | |
|   Mixed lobules | 18 (90.0) | 12 (60.0) | |
|   Predominantly Type 1, no Type 3 | 2 (10.0) | 8 (40.0) | |